# Unconventional non-uniform local lattice distortion in dilute Ti-Mo solid solution

Qing-Miao Hu[1,2] & Rui Yang[1]

**The substitutional solute atom induced local lattice distortion (LLD) in dilute metal solid solution was believed to be uniform that may even be modeled by using soap bubble raft. Contrary to this conventional picture, we report in this paper that the Mo induced LLD in dilute Ti-Mo solid solution is highly non-uniform as evidenced by our first principles calculations. The non-uniform LLD is ascribed to the Jahn-Teller splitting of the degenerated *d* states of Mo atom. We propose that the substitutional solid solutions with non-uniform LLD should satisfy two conditions. With which, the solid-solutions suffering from non-uniform LLD are predicted. The non-uniform LLD is expected to result in non-spherical stress field around the solute atom, and, therefore, challenges the application of classical solid solution hardening model to this kind of solid solutions.**

Solid solution hardening (SSH) is a widely employed approach to improve the mechanical properties of metals. To evaluate the SSH effect, about 80 years ago, classical SSH model was initiated by Mott and Nabarro[1] and developed further by many other researchers[2-6] later on. The model is still in use up to date[7-12] because it is much simpler and less costly than the state-of-the-art, more sophisticated technologies[13-16]. The classical SSH model for substitutional solid solutions was built by considering the interaction between the solute atoms and the dislocation within the framework of elasticity theory. The solute atom induced local stress field was treated as being spherical. This is based on the assumption that the solute and host atoms are elastic misfit spheres, and the distances between the solute atom and its nearby host atoms shrink or dilate uniformly depending on the solute-host atomic size misfit. With which, the misfit strain may be defined simply as $\varepsilon = (d - d_0)/d_0$ with $d$ and $d_0$ respectively being the solute-host and host-host interatomic distance. The misfit strain enters into the SSH model as one of the key parameters. According to Cottrell's model[2], the interaction energy and force between solute atom

---

[1] Institute of Metal Research, Chinese Academy of Sciences, Wenhua Road 72, Shenyang 110016, China
[2] Email: qmhu@imr.ac.cn

and dislocation and the consequent SSH effect are proportional to the misfit strain.

The aforementioned assumption works fine in general for substitutional metal solid solutions because the atomic bond in metal is non-directional. For such systems, the solute induced local lattice distortion (LLD) may even be modeled by using the soap bubble raft as researchers did in early days when no accurate computational tools were available. The development of the modern computation technologies such as the first-principles methods based on density functional theory makes the accurate calculation of LLD feasible. By using these methods, the uniform LLDs have been reported for many dilute solid solutions such as face centered cubic (fcc) Al-[9,17], Cu-[17,18], Ni-[17-19], body centered cubic (bcc) Fe-, Nb-, V-[18], Li-[20], and hexagonal close packed (hcp) Ti-[21,22] based ones. However, our present work demonstrates that, for some of the substitutional metal solid solutions, the solute induced LLD is highly non-uniform, which put the application of the classical SSH model to this kind of solid solutions into question.

In the present work, we first calculate the LLD induced by a substitutional Mo atom in hcp Ti matrix in the conventional way[9,19-22] by using the first-principles plane wave method implemented in Vienna Ab initio Simulation Package (VASP)[23]. A supercell Ti is constructed for pure Ti. One of the Ti atom is then replaced by a Mo atom. Both the pure Ti and Ti-Mo supercells are fully relaxed by minimizing the external stresses and interatomic forces.

Fig. 1 shows the cuboctahedrons cut from the relaxed pure Ti (Fig. 1a) and Ti-Mo (Fig. 1b) supercells. As seen in Fig. 1a, for the pure Ti supercell, the distance between the nearest neighboring Ti atoms in the same layer of the (0001) basal plane (denoted as in-plane Ti atoms) is 2.934 Å while those between the Ti atoms in different layers of basal planes (denoted as out-plane Ti atoms) is 2.875 Å, in good agreement with the experimental values (2.951 Å and 2.894 Å, respectively) [24]. For the Ti-Mo supercell, the relaxed distances between the Mo and 6 in-plane Ti atoms are 2.913 Å while those between the Mo and 6 out-plane Ti atoms are 2.863 Å. The solute Mo atom results in the in-plane and out-plane nearest neighboring distances shrinking by 0.72\% and 0.42\%, respectively. Namely, the Mo induced LLD in hcp Ti calculated in the conventional way is indeed uniform, consistent with the elastic misfit sphere assumption. The Mo locates exactly at the highly symmetric hcp lattice site (the center of the cuboctahedron) after the geometric optimization.

The dynamic stability of the above optimized Ti-Mo system is evaluated by calculating the lattice vibration frequencies. Fig. 2a displays the phonon dispersion. Surprisingly, there exist two significantly low-lying phonon bands with imaginary frequencies of about -2.54 THz, corresponding to the vibrations of the Mo atom along the **a** and **b** lattice vectors in real space. Imaginary frequencies also appear near the G (0.0, 0.0, 0.0) point for another band corresponding to the vibration of Mo along the **c** direction. The imaginary frequencies mean that the configuration with Mo occupying the high symmetry substitutional position is actually dynamically unstable. The reason that the Mo atom keeps sitting at the high symmetry lattice site is that, due to the symmetry of the lattice, the interatomic forces acting on the Mo atom are zero in all directions. Therefore, the position of Mo is not updated during the geometric optimization.

Considering the imaginary frequencies, we move the Mo atom slightly away from the high symmetry position so as to break the symmetry of the lattice, and optimize the supercell again. This results in Mo-centered optimized cuboctahedron as shown in Fig. 1c. We see that the cuboctahedron is highly distorted. Two in-plane nearest neighboring Mo-Ti distances shrink by 8.1% while the other four dilate by 3.0~4.6%. Two out-plane nearest neighboring Mo-Ti distances dilate by 7.1% and the other four shrink by 3.3%. The Mo atom moves obviously away from the high symmetry position. The LLD around Mo is now highly non-uniform.

The dynamic stability of the low symmetry configuration with non-uniform LLD is examined as well. Fig. 2b shows the corresponding phonon dispersion. The low-lying phonon bands for the vibrations of Mo along the **a** and **b** directions move up and the imaginary frequencies become positive, except for a few negligible negative frequencies (> -0.1 THz) near G (0.0, 0.0, 0.0). For the vibration of Mo along **c** direction, the imaginary frequencies completely disappear. The phonon dispersion shown in Fig. 2b demonstrates that the low symmetry position is dynamically stable for Mo to occupy.

The dynamic instability/stability are in accordance with the energetics of the Ti-Mo systems with uniform and non-uniform LLDs. The total energy of the optimized Ti-Mo supercell with uniform LLD is about -288.386 eV whereas that of the supercell with non-uniform LLD is 0.130 eV lower.

The above results demonstrate that, contrary to the conventional picture about the LLD in substitutional metal solid solutions, the LLD induced by solute Mo atom in Ti-Mo system is highly non-uniform. The non-uniform LLD should lead to stress field deviating from spherical. This means that the classical SSH model based on the elastic misfit sphere and uniform LLD assumption might not work properly for Ti-Mo solid solution any more. On the other hand, as described previously in this paper, the non-uniform LLD is about one order of magnitude larger than the uniform one. Consequently, the actual SSH effect is expected to be much stronger than that predicted by using the classical SSH model.

To explain why the low symmetry position is more stable than the high symmetry one for Mo to occupy, we examine the electronic structures of both configurations. Fig. 3 compares the electronic density of states (DOS) of Mo atoms occupying the high and low symmetry positions. As seen in the figure, the *d* states of the high symmetry Mo atom possess a high peak right at the Fermi level. Relatively lower peaks exist at about -1.0 eV and 1.0 eV. With Mo moving to the low symmetry site, the DOS at Fermi level is significantly lowered. Two small peaks appear at energies slightly lower and higher than the Fermi level. The DOS peak at -1.0 eV rises whereas the peak at 1.0 eV becomes lower and moves to higher energy (1.8 eV).

The difference between the DOSs of Mo atoms at high and low symmetry positions may be ascribed to the Jahn-Teller splitting effect[25]. At the high symmetry position, the Mo atom subjects to highly symmetric cuboctahedron crystal field formed by the surrounding Ti atoms such that some of the *d* states of Mo are degenerated at the Fermi level. With the Mo atom moving to the low symmetry position, the cuboctahedron and the related crystal field becomes significantly distorted, making the degenerated *d* states split. The Jahn-Teller splitting lowers the total energy of the system and leads to a more stable low symmetric occupation of Mo.

Fig. 4 shows the bonding charge density (differences between the charge density of the converged supercell and the superposition of free atoms) for the (0001) planes of Ti-Mo systems with Mo occupying respectively the high and low symmetry positions. For the high symmetry case, the bonding charge distributes homogenously among Mo and its surrounding Ti atoms, indicating that the bonds between Mo and Ti atoms remain non-directional. However, for the low symmetric case, the bonding charge among Ti1 and Ti2 and Mo increase evidently, i.e., the bonding between Mo and Ti1 and Ti2 becomes stronger. The bonding between the Mo and Ti atoms is no longer non-directional.

A certain concern about this work is whether the non-uniform LLD is universal and occurs in some other metal solid solutions or not. According to the physics underlying the non-uniform LLD as discussed above, we may speculate the conditions required for the occurrence of the non-uniform LLD.

First, the crystal lattice of the host element should not favor high degeneracy of $d$ orbitals at the Fermi level. For the transition metals (TM) with hcp structure, such as Ti, Zr, and Hf early in the Chemical Element Periodic Table (CEPT), the DOS shows a pseudogap right at the Fermi level, indicating that the hcp lattice of these kind of TM metals is unfavorable to accommodate atoms with degenerated $d$ orbitals. Therefore, the TM metals with hcp structure are the ideal hosts for the solute atoms to induce non-uniform LLD. In general, pseudogaps do not exist at the Fermi levels of the DOSs of bcc TM (e.g., Fe, V, Nb) in the middle of CEPT and for the late fcc TM (e.g., Ni) and nobel metals (e.g., Cu). Namely, the bcc and fcc lattices of these metals might be able to accommodate atoms with degenerated $d$ orbitials. Thus, the solid solution with bcc and fcc TM and nobel metal hosts may not suffer from the non-uniform LLD.

Second, the solute atoms should have sufficient occupied $d$ orbitals that may degenerate at the Fermi level, facilitating the Jahn-Teller splitting to take effect. This is the characteristic of TM elements in the middle of the CEPT such as Mo involved in the present work. The $d$ orbitals of the noble metal elements (e.g., Cu, and Zn) are almost fully occupied and well below Fermi level.

Consequently, non-uniform LLD might not show up in solid solutions with solutes of nobel metal atoms due to the absence of Jahn-Teller splitting.

The solid solutions satisfying the above conditions are the ones with early TM elements (e.g, Ti, Zr, and Hf) with hcp structure as host and with the middle TM elements (e.g., Cr, Mn, Fe, Mo, Tc, Ru, W, Re, Os) as solute. In these solid solutions, the solute induced non-uniform LLD may occur. Further work is encouraged to verify the prediction.

In summary, our first-principles calculations demonstrated that the substitutional Mo atom moves significantly away from the high symmetry hcp lattice site in dilute Ti-Mo solid solution. The induced local lattice distortion is highly non-uniform. This finding challenges the application of the classical solid solution hardening model to the related solute solutions because the classical model is based on the assumption that the substitutional atoms induced local lattice distortion is uniform and the corresponding stress field around the solute is spherical. The non-uniform local lattice distortion is about one order of magnitude stronger than the uniform one induced by Mo atom locating at the high symmetry site. Consequently, for the solid solutions undergoing non-uniform local lattice distortion, the classical solid solution hardening model might significantly underestimate the actual hardening effect. The physics underlying the non-uniform local lattice distortion is the Jahn-Teller splitting of the degenerated *d* orbitals of the solute atom. Considering the underlying physics, we proposed that the solid solutions with non-uniform local lattice distortion should satisfy two conditions. Based on which, we predicted that the non-local lattice distortion may occur in solid solutions with early transition metal host and middle transition metal solute.

**Methods**

In our first-principles calculations, the supercell size of the pure Ti and Ti-Mo systems is 3×3×2 times of the hcp unit cell. The project augmented wave potential[26] is employed for the interaction between the valence electrons and ionic cores. The generalized gradient approximation parameterized by Perdew, Burke, and Ernzerhof[27] is adopted to describe the electronic exchange and correlation. The plane-wave cutoff energy is set as 450 eV. A *k* point mesh of 4×4×4 is used to sample the Brillouin zone. The geometric structures of the Ti and Ti-Mo supercells are fully relaxed by using the first-principles plane-wave pseudopotential method. The energy tolerance for the electronic minimization is set as $1\times10^{-6}$ eV/atom and the force tolerance for the geometric optimization is set as $1\times10^{-2}$ eV/atom eV/Å. For the phonon dispersion calculations, we adopted the small displacement approach[28]. The symmetrized force constant is calculated with atomic displacements of $\pm0.015$ Å.

25. Sturge, M.D. The Jahn-Teller effect in solids. *Solid State Phys.* **20**, 91-211 (1968).
26. Blöchl, P.E. Projected augmented-wave method. *Phys. Rev. B* **50**, 17953-17979 (1994).
27. Perdew, J.P., Burke, K. & Ernzerhof, M. Generalized gradient approximation made simple. *Phys. Rev. Lett.* **77**, 3865-3868(1996).
28. Alfè D. PHON: A program to calculate phonons using the small displacement method. *Comput. Phys. Comm.* **180**, 2622-2633 (2009).
**Acknowledgements**

The authors are very grateful to Prof. Dario Alfè from University College London, UK, for the help in our calculations of the phonon dispersion. This work is financially supported by Natural Science Foundation of China under grant No. 91860107 and the National Key Research and Development Program of China under grant No. 2016YFB07013019

**Figures**

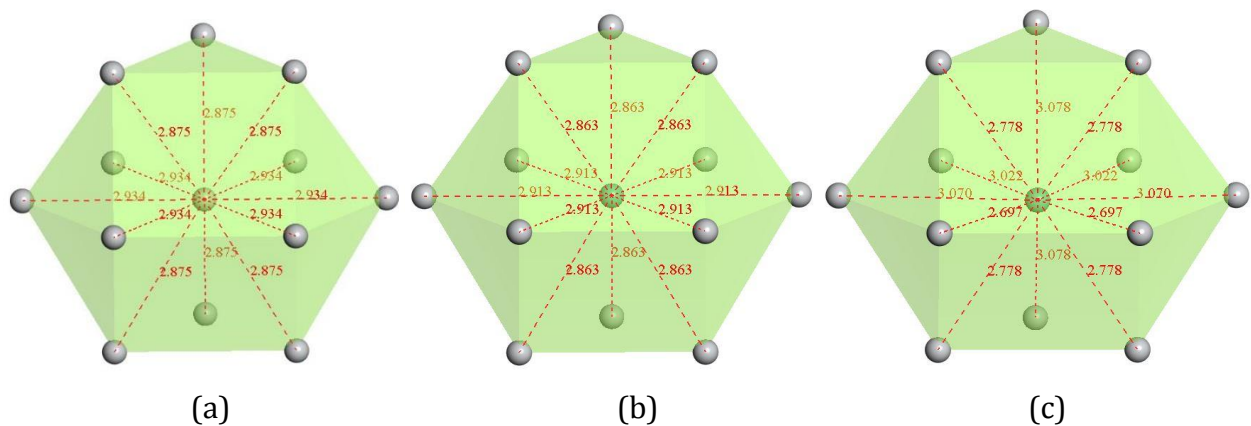

Fig. 1 Cuboctahedrons cut from the optimized supercells of pure Ti (a), uniformly distorted Ti-Mo supercell (b), and non-uniformly distorted Ti-Mo supercell (c). For the Ti-Mo supercells, the Mo atom locates at the center of the cuboctahedrons. The middle seven atoms are in the same layer of the (0001) basal plane of the hexagonal close packed structure while the top three and bottom three atoms are in the (0001) layers next to the middle one.

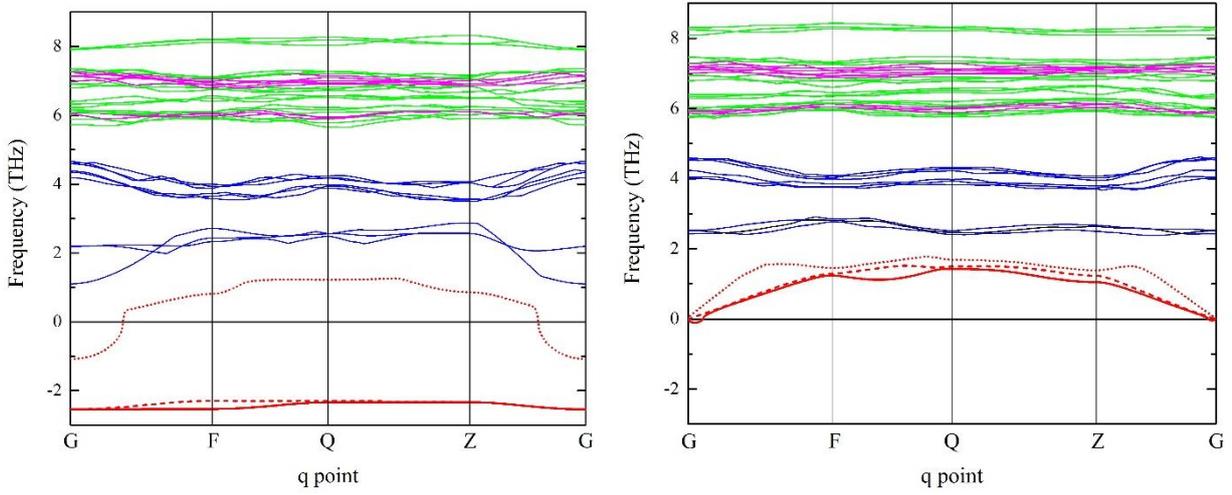

Fig. 2 Phonon dispersions of the Ti-Mo systems with uniform (a) and non-uniform (b) local lattice distortions. The phonon dispersion is along the Brillouin zone q point path: G (0.0, 0.0, 0.0) → F (0.0, 0.5, 0.0) → Q (0.0, 0.5, 0.5) → Z (0.0, 0.0, 0.5) → G (0.0, 0.0, 0.0). To avoid crowdedness in the figure, only the vibrational frequencies of Mo (red) and its nearest neighboring in-plane (green) and out-plane (magenta for top and blue for bottom, see Fig. 1) Ti atoms are shown. The red solid, dashed, and dotted curves represent respectively the phonon bands for the vibrations of Mo along the **a**, **b**, and **c** directions.

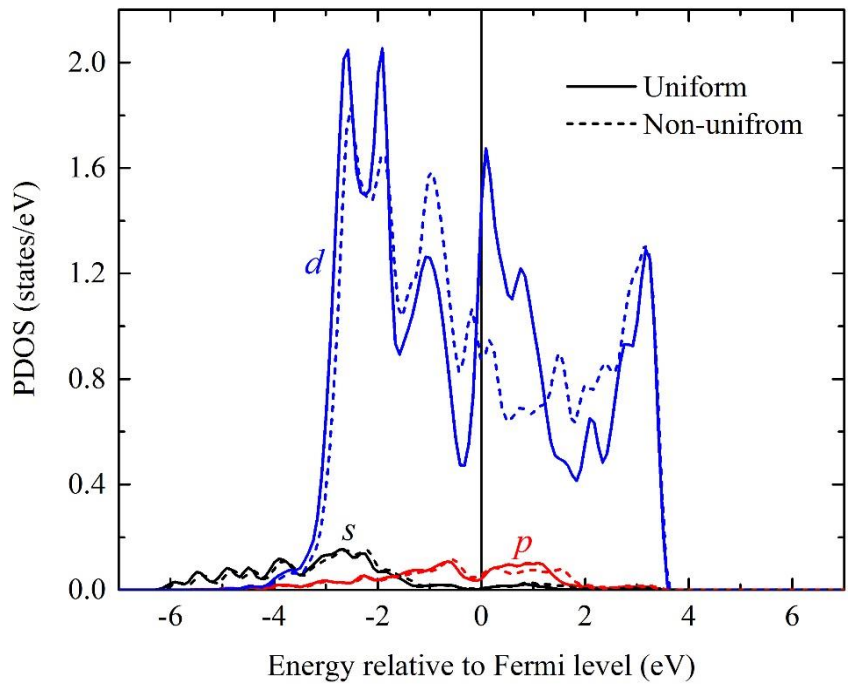

Fig. 3 Electronic density of states of Mo in Ti supercells with uniform (solid lines) and non-uniform (dash lines) local lattice distortions.

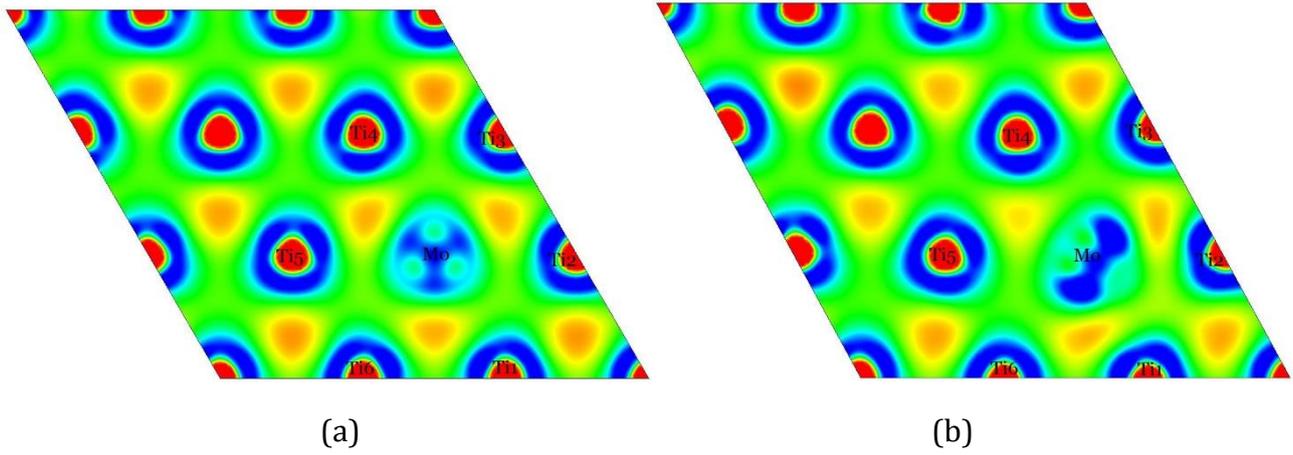

Fig. 4 Bonding charge density of the Mo containing (0001) planes of the Ti-Mo supercells with uniform (a) and non-uniform (b) local lattice distortions, scaled from -0.02 to 0.02 e/Å$^3$ with color from blue to red.